\documentclass[10pt]{article}
\usepackage{hicss}
\usepackage{times}
\usepackage[none]{hyphenat}
\usepackage{url}
\usepackage{latexsym}
\usepackage{indentfirst}
\usepackage{graphicx}
\usepackage{array}
\usepackage[hidelinks]{hyperref}
\usepackage{color}
\hypersetup{breaklinks=true}
\newcolumntype{P}[1]{>{\centering\arraybackslash}p{#1}}
\newcolumntype{M}[1]{>{\centering\arraybackslash}m{#1}}

\usepackage{amsmath,amssymb,amsfonts,amsthm}
\newtheorem{definition}{Definition}

\title{Constraint-based recommender system for crisis management simulations}

\author{Luyen Le Ngoc$^{1,2}$\\
    {\underline{ngoc-luyen.le@hds.utc.fr}} \\\And
  Jinfeng Zhong$^3$\\
  {\underline{jinfeng.zhong@dauphine.eu} }\\\And
  Elsa Negre$^3$\\
  {elsa.negre@dauphine.fr}\\\And
    Marie-Hélène Abel$^1$\\
  {marie-helene.abel@hds.utc.fr}
  \AND 
  \hspace{-1.4cm}1: Université de technologie de Compiègne, CNRS, Heudiasyc, CS 60319 - 60203 Compiègne Cedex, France.
 \\
 \hspace{-7.9cm}2: Vivocaz, 8 B Rue de la Gare, 02200, Mercin-et-Vaux, France.
 \\
  \hspace{-2.2cm}3: Paris-Dauphine University, PSL Research Universities, CNRS UMR 7243, LAMSADE, Paris, France.\\
  }

\date{}

\begin{document}
\maketitle
\begin{abstract}
In the context of the evacuation of populations, some citizens/volunteers may want and be able to participate in the evacuation of populations in difficulty by coming to lend a hand to emergency/evacuation vehicles with their own vehicles. One way of framing these impulses of solidarity would be to be able to list in real-time the citizens/volunteers available with their vehicles (land, sea, air, etc.), to be able to geolocate them according to the risk areas to be evacuated, and adding them to the evacuation/rescue vehicles. Because it is difficult to propose an effective real-time operational system on the field in a real crisis situation, in this work, we propose to add a module for recommending driver/vehicle pairs (with their specificities) to a system of crisis management simulation. To do that, we chose to model and develop an ontology-supported constraint-based recommender system for crisis management simulations.

\end{abstract}

\section{Introduction}

Climate change is a hot topic that impacts and involves everyone. These rapid and consequential changes are largely due to increasing (and sometimes harmful) human activity. Unfortunately, ``the effects of human-caused global warming are happening now, are irreversible on the timescale of people alive today, and will worsen in the decades to come''\footnote{https://climate.nasa.gov/effects/}. Thus, these changes/disturbances threaten the environmental balance and cause an increase in natural risks such as hurricanes, fires or rapid floods. Natural disasters are a constant cause of human suffering and economic loss around the world. Crisis management stakeholders must therefore react effectively to these disasters: (i) before, to anticipate risks, reduce them and prepare populations, (ii) during, to react and put in place actions (such as sheltering populations) and, finally (iii) after, to consider lived experience, predict and reduce the impact of similar future events. The material and economic stakes are important, but the human stakes are even more so. Thus, one of the major issues in crisis management is the safety of populations, which often results in the evacuation of populations from risk areas. For this, it is necessary to know the area considered to be at risk in terms of geography but also of people. Indeed, in the event of a disaster, some people are able to shelter themselves but others are not. The latter are vulnerable and therefore need help, such as the disabled, the elderly, the injured, etc. This help (often) results in the deployment of emergency/rescue vehicles to evacuate vulnerable populations located in risk areas.

Today, there are several techniques to help rescue services improve the sheltering process, including optimization and simulation. Optimization helps create effective evacuation plans, and simulation helps explore what-if scenarios. Coupled with the use of geospatial data to visualize the dynamics and display useful measures in a realistic environment, it is thus possible to have crisis management tools that are practical, effective for decision-making, and reusable during different disasters \cite{laatabi2022coupling}.

The existing models on which these crisis management tools are based are generally limited to self-evacuation or evacuation by emergency vehicles \cite{laatabi2022coupling}. However, today, the participation of populations is at the heart of crisis management and many citizens want to be active in their environment. And, although there are entities such as civil protection, many impulses of solidarity and mutual aid are manifested in a crisis situation by ``simple'' citizens/neighbours. In fact, ``natural disasters know no borders – neither do solidarity and mutual support in times of need'' \cite{jcm21}. The resulting actions are sometimes ad-hoc and need to be at least coordinated to be optimized and in agreement/correlation with the action plans defined by public decision-makers.

In the context of the evacuation of populations, some citizens/volunteers may want and be able to participate in the evacuation of populations in difficulty by coming to lend a hand to emergency/evacuation vehicles with their own vehicles. Indeed, the owner of a 9-seater mini-van could evacuate 8 additional people and thus increase the capacity of the evacuation vehicles, or the owner of a 6-seater boat could help evacuate 5 people during a flood.

One way of framing these impulses of solidarity would be to be able to list in real-time the citizens/volunteers available with their vehicles (land, sea, air, etc.), to be able to geolocate them according to the risk areas to be evacuated, and adding them to the evacuation/rescue vehicles. If more vehicles with drivers are available to evacuate populations then it is optimistic to think that more people can be sheltered. 

Because it is difficult to propose an effective real-time operational system on the field in a real crisis situation, in this work, we propose to add a module for recommending driver/vehicle  pairs (with their specificities) to a system of crisis management simulation (like \cite{laatabi2022coupling} which anticipates sheltering populations in crises). The idea is therefore to add alternatives for the what-if scenarios and hope to evacuate more people in less time in the best possible conditions. Technically, our proposal should be a brick of a resource allocation system in the event of a crisis which should support the design principles and specifications of a Dynamic Emergency Response Management Information System (DERMIS) \cite{dermis}.

The remainder of this paper is as follows. The following section presents some related works about ontologies (to consider driver/vehicle specificities) and recommender systems, especially in the context of crisis management. The third section details our proposal of a constraint-based recommender system for crisis management simulations. In the fourth section, we present our prototype and its application on a detailed use case. Finally, we conclude and propose some future works in the last section.

\section{Related works}
In this section, we will introduce the related works from which we have borrowed concepts.


\subsection{Ontologies}
Crisis management is a multidisciplinary and complex domain and 
requires flexibility and improvisation in different situations from crisis actors/volunteers involved. They have different backgrounds and thus need to have a common vocabulary to communicate and understand each other. 
Semantic modeling using ontology is suggested to standardize the vocabularies and concepts of the crisis management domain in order to reduce complexity and avoid misunderstandings of terminologies used in different stages of management and intervention activities \cite{mughal2021orffm}.

By structuring and organizing a set of terms or concepts within a domain in a hierarchical way and by modeling the relationships between these sets of terms or concepts using a relation descriptor, an ontology provides a method for knowledge representation that is able to describe and interpret semantic information used for the communication between different systems, among human beings, between human beings and systems \cite{rodriguez2003determining}.  An ontology allows identifying concepts, taxonomies, relations, and rules for defining and representing knowledge in a particular domain. Therefore, ontology facilitates building knowledge base models for various tasks such as information modeling, knowledge engineering, sharing, or data linking \cite{jakus2013concepts}.

In the context of knowledge sharing, an ontology is a formal and explicit description of shared knowledge that consists of a set of concepts in a domain and the relationships between those concepts \cite{guarino1995towards}. Therefore, ontologies can be employed for various practical purposes: (1) the structure and organization of domain information: an ontology is built on the basis of the natural structures of information by making it possible to visualize the concepts and their relations, and to reuse domain knowledge,  (2) semantic search improvements: instead of searching by keywords, ontology search can return synonyms from query terms, (3) the integration of data from different sources, different languages \cite{abiteboul_2011}. 
Alternatively, an ontology can be organized and stored on NoSQL databases such as MongoDB or Neo4j that gain advantages in terms of integration and performance in information systems.

The design and use of ontologies in crisis management systems can help to define logical semantic rules for organizing information and supporting a decision-making process. Indeed, various ontologies have been proposed and developed for modeling different aspects and situations in crisis management. In \cite{babitski2009ontology}, the authors developed an ontology that organizes and integrates heterogeneous information for resource and damage descriptions in disaster management. In \cite{de2005siadex}, the authors  constructed an ontology aimed at planning objects and activities related to the forest fighting plan. In \cite{benaben2008metamodel}, the authors described their work on \textit{ISyCri} ontology that includes different categories of entities affected by crises such as goods, natural sites, people, and civil society. And more specifically, their crisis ontology models can elaborate and deduct a crisis solving collaborative process through ontological reasoning and rules. Or in  \cite{zhong2017geo}, the authors developed ontologies to represent the geospatial characteristics and events used in a meteorological disaster system that is able to provide semantic integration for computer-aided decisions in emergency management, including routine and urgent activities such as resource planning and prediction of the next disaster. In \cite{gaur2019empathi}, the authors designed \textit{empathi} ontology that conceptualizes and organizes situational and environmental awareness subjected to hazards. Their ontology aids in crisis management, hazard situational awareness, and events used in emergency situations. In summary, current existing ontologies for the domain of crisis management are completely dependent on each specific crisis scenario with different knowledge and information about the kind of crisis and/or area. These ontologies lack systematical analysis of the model and organization for managing and allocating civil/volunteer driver/vehicle resources to support evacuation activities in crisis management. 

Recommender systems based on a knowledge base represented by means of ontologies have been widely developed for various domains, including e-commerce \cite{le2022towards}, education, and engineering \cite{abel2021memorae}, remarkably, in emerging domain crisis management \cite{jain2018ontology, mehla2020ontology}. We detail in the next section, recommender systems and their applications for suggesting relevant resources in crisis management.

\subsection{Recommender systems} \label{sec:recommender_system}

Recommender systems are special types of decision support tools that help users find the appropriate items for users through information retrieval. Collaborative filtering approaches and content-based approaches are the two types of recommender systems widely used in various domains. Content-based approaches explore items' attributes such as categories and keywords. Both collaborative filtering approaches and content-based approaches have achieved great success in many domains. However, both methods suffer from data sparsity: when the interactions between users and items are sparse, the prediction of ratings becomes less accurate, which is known as the cold start problem \cite{lam2008addressing}. This is the case in crisis management where recommender systems help decision-makers to adopt the best actions \cite{negre2013towards} or distribute available resources \cite{kouzmin2008crisis}, which means that collaborative filtering approaches are not appropriate. As for content-based approaches, they model items' attributes. Therefore, they are more adapted for content recommendations such as movies and music. And these approaches also require users' ratings towards items, in crisis management, such data are usually not available.

Knowledge-based recommender systems can be applied to alleviate the challenges mentioned above. The advantage of such recommender systems is that no rating data is needed for computing recommendations \cite{aggarwal2016knowledge}. There are two types of basic knowledge-based recommender systems: namely constraint-based recommender systems \cite{felfernig2015constraint} and case-based recommender systems \cite{bridge2005case}. Both approaches require that users should specify their demands and the system returns recommendations, which is the case in crisis management. Decision-makers should indicate the demands during a crisis and the system tries to compute appropriate recommendations. The difference lies in the way how knowledge is applied. Case-based recommender systems find similar items by applying different types of similarity (e.g. semantic similarity) while constraint-based recommender systems compute recommendations that satisfy a set of explicit predefined constraints (rules). Indeed, knowledge-based recommender systems have been applied in crisis management to help decision-makers adopt the best actions. In \cite{jain2018ontology}, the authors design a recommender system that can perform case-based reasoning and rule-based reasoning supported by ontologies. This system can generate a list of actions that help to reduce damages caused by earthquakes. In \cite{mehla2020ontology}, the authors present an ontology-supported recommender system that helps decision-makers to respond effectively during emergencies, recommendations are generated by combining case-based and rule-based reasoning like in \cite{jain2018ontology}. In \cite{zhang2016ontology}, the authors applied rule-based reasoning using Semantic Web Rule Language (SWRL) to help manage crises caused by meteorological disasters.

 During a crisis, some constraints must be satisfied rather than just being similar. For example, when evacuating a population during a flood, if the place is not accessible by ground vehicles then boats or helicopters must be utilized. In this case, constraint-based recommender systems are more appropriate since they help to filter items that satisfy pre-defined constraints. We wish to propose a recommender system that can help evacuate the population during a crisis. In \cite{wex2014emergency, chen2011collaborative}, the authors propose frameworks that help to allocate resources with constraints being satisfied. Nonetheless, the authors did not base their methods on ontology, limiting the reusability of their methods. These works \cite{jain2018ontology,mehla2020ontology,zhang2016ontology}, all present a recommender system to help decision-makers better distribute public resources, such as ambulances, fire trucks, etc. However, we argue that public resources may sometimes be limited and the resources may not be well located. During a crisis, public resources may not be sufficient and in this case, other resources have to be explored. By ``not well located'', we mean that sometimes, the public resources such as ambulances could be too far to be reached. On the other hand, civil/volunteer resources such as cars, and buses owned by private companies are scattered, which makes them more easily accessible. Besides lacking public human resources is another vital issue during a crisis. Civil cars are usually associated with drivers, which may greatly help to ease the shortage of public human resources. In the next section, we will present in detail our ontology-supported constraint-based recommender system for crisis management simulations.

\section{Our proposition}
\vspace{-0,2cm}
In this section, we first lay out our proposition: an ontology-supported constraint-based recommender system that helps to distribute pairs of driver/vehicle during a crisis simulation. 

\subsection{Problem formulation}
\vspace{-0,3cm}
In the context of crisis management, the organization and use of civil/volunteer resources including drivers and their vehicles in evacuation activities allow adding more solutions that can be used to move all affected populations to safe locations, besides using the public resources of evacuation centers which can be limited in both human and material resource \cite{laatabi2022coupling}. Mobilizing and allocating pairs of civil/volunteer driver/vehicle resources requires calculation and optimization in terms of estimated response time between the vehicle's location and the rescue point, as well as the number of required pairs of civil/volunteer driver/vehicle resources. This research attempts to model and propose a system that manages these pairs of civil/volunteer driver/vehicle resources and suggests relevant allocating resources for evacuating all affected populations by the crisis. Therefore, the two key problems are: ($P1$) to organize data and information involved in pairs of civil/volunteer driver/vehicle resources and ($P2$) to recommend optimal solutions under the constraints of capacity of pairs of driver/vehicle resources,  response time, and context. The first one ($P1$) focuses on choosing the most relevant model used for organizing information and data, with storage and exploitation of useful information in the most appropriate way in a crisis management context. The second one ($P2$) is a matter of designing and developing a recommender system that can propose solutions to resource allocation in a suitable situation.

\begin{definition}{Ontological modeling task} is the process of engineering different components of ontologies including a set of concepts, a set of relations between concepts, a set of attributes, a set of rules, and restrictions in order to construct a knowledge base for a particular usage or domain.
\end{definition}

For the problem $P1$, the ontological modeling task can help to construct a crisis management knowledge base that can capture and represent concepts, and relations of the domain knowledge, which can be enriched by adding individuals that represent instances.  Thus, ontological modeling is applicable in providing structured information on resources, locations, and actors/volunteers in crisis management.

Knowledge-based recommender technologies by means of ontology can help to tackle the problem $P2$ by exploiting explicit requirements of rescue points about the number of people, priority level, and knowledge about the context and available resources for the calculation of relevant solution recommendations.

\begin{definition}{A constraint-based recommender system for the resource allocation} is defined by using 4 sets:  the set of moving resources $\mathcal{R}$, the set of rescue points and their requirements $\mathcal{P}$, the set of shelters $\mathcal{S}$, and the set of constraints $\mathcal{C}$. A relevant solution recommendation is calculated based on the concrete element of sets $\mathcal{R}$, $\mathcal{P}$, and $\mathcal{S}$ such that the specified constraints $\mathcal{C}$ are satisfied.\end{definition}

A moving resource is a pair of civil/volunteer driver/vehicle that is used as a complete resource.  For example, we illustrate the simplified information about the sets in crisis management: \vspace{-0,2cm}
\begin{itemize}
	\item $\mathcal{R}$ = \{(Josept Brault/Minibus Peugeot, 15 seats),  (Julien Laisne/SUV X5 BMW, 5 seats), ...\}, 
	\item $\mathcal{P}$ = \{(Rescue Point 01, 100 affected people, priority 1), (Rescue Point 02, 72 affected people, priority 2), ...\}, 
	\item $\mathcal{S}$ = \{(Rose Gymnasium, Compiègne), (Saint German Gymnasium, Compiègne), ...\}, 
	\item $\mathcal{C}$ = \{Minimum the response time, Maximum number of evacuated people, ...\}. 
\end{itemize}

\begin{definition}{Recommendation task for allocating moving resources}
	is defined as a constraint satisfaction problem $(\mathcal{R}, \mathcal{P}, \mathcal{S}, \mathcal{C})$ based on the assignment and calculation of the number of moving resources in the set $\mathcal{R}$ allocated for a rescue point in $\mathcal{P}$ such that it satisfies and does not violate any of the constraints in $\mathcal{C}$.
\end{definition}

An optimal solution recommendation for allocating moving resources in $\mathcal{R}$ will propose a list of available moving resources used for transferring the evacuees from rescue points $\mathcal{P}$ to shelters $\mathcal{S}$ with a minimum travel time. To dive into the work, we will detail the development of an ontology in crisis management for resources and related factors in the next section.

\subsection{Construction of ontology}\label{sec:construction_of_ontology}

\begin{figure*}
	\centering
	\includegraphics[width=0.9\linewidth]{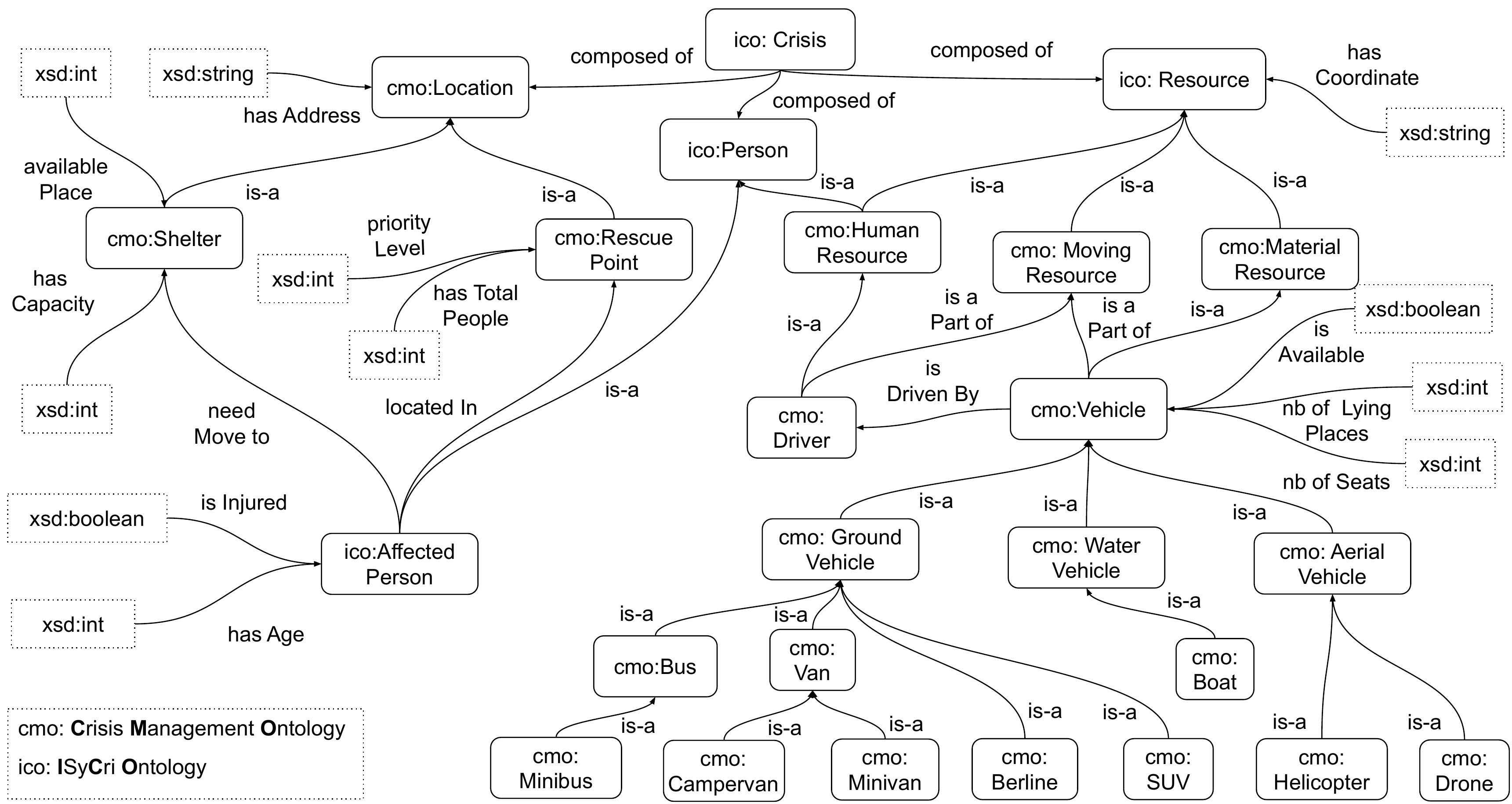}
	\caption{An ontology about driver/vehicle resources, locations and related people in a crisis management}
	\label{fig_01}
\end{figure*}

Thorough planning and preparedness are a prerequisite for participating and responding in emergency situations. The preparedness is demonstrated through organizing resources and planning their use, and allocating these resources to each disaster-affected site. Crisis management modeling in these situations helps to indicate an overview of the factors that provide necessary information for calculating acceptable solutions and aiding decision-making. The structure and organization of information and data sources by using ontologies allow taking advantage of the ontological modeling's strengths, including the capacities to represent explicit concepts and connect them through relationships, to facilitate the sharing and reuse of knowledge between people and widely distributed systems in a real-time way. An ontology owns a reasoning mechanism by using a reasoner that allows deducting additional knowledge. Furthermore, it is possible to declare inference rules to ensure the consistency of description or extract information that is not explicit. 

The main objective of the construction of ontology is to provide a standard model with terminology and vocabulary for gathering information about available resources (e.g, vehicles, drivers) that an organization (e.g, Municipal Council) will mobilize in order to evacuate affected people during a crisis. The core concepts of the ontology are defined based on the important entities with their characteristics in the context of crisis management. Therefore, we construct our ontology based on the inheritance of \textit{ISyCri} Ontology in \cite{benaben2008metamodel} by using concepts related to the description of crisis, affected people, and resources. More specifically as illustrated in Figure~\ref{fig_01}, we adapt and develop our ontological model around three main entities: resources, people, and locations. Firstly, the resources are distinguished into human resources, material resources and moving resources.  In our case, human resources are citizens/volunteers who participate in rescue and evacuation operations. While material resources include categories of vehicles and their description information. A moving resource will be configured by a default association of a vehicle and a driver (i.e. a pair of civil/volunteer driver/vehicle resource). Conventionally, a moving resource $R$ is defined as a set of triplets as follows:\vspace{-0,2cm}
\begin{equation}
	R = \{a^R_1, a^R_2, ..., a^R_k\}
\end{equation} where $a^R_i$ presents the triplet $a^R_i$ $=\langle subject_i,$ $predicate_i,$ $object_i \rangle$. In other words, the triplet $a^R_i$ can also be denoted as  $\langle resource_i,$ $propertie_i,$ $statement_i \rangle$. For example, we have the description of a moving resource: ``\textit{A Toyota Sienna of 8 seats is driven by Henri Le who lives at 5 Place de Bretagne, Brest}'', this information can be expressed in a set of triplets as $R \: = \: \{ \langle Henri\_Le/Toyota\_Sienna, \: rdf:type, \: cmo:MovingResource \rangle, \quad \langle Toyota\_Sienna, \:\:\:\: rdf:type, \: 
\\cmo:Minivan \rangle, \:\:\: \langle Toyota\_Sienna, \:\:\: is\_a\_Part\_of, \: 
\\Henri\_Le/Toyota\_Sienna \rangle, \: \langle Henri\_Le, \: rdf:type, \:
\\cmo:Driver \rangle,\: \langle Henri\_Le, \: is\_a\_Part\_of, \: Henri\_Le
\\/Toyota\_Sienna \rangle, \:\:\: \langle Toyota\_Sienna,\:\:\: nb\_of\_Seat, \: 
\\8\_places \rangle \}$. Secondly, the identification and organization of locations play an extremely important role in our case. Each location needs to be specifically identified with information about their address. In general, the locations are separated into rescue points and shelters. Rescue points are sites where affected populations aggregate and are transited to a shelter by a moving resource (or a pair of civil/volunteer driver/vehicle resource). A rescue point $P$ can be defined as a set of triplets as follows:\vspace{-0,2cm}
\begin{equation}
	P = \{a^P_1, a^P_2, ..., a^P_n\}
\end{equation} where $a^P_j$ denotes the triplet $a^P_j$ $=\langle subject_j,$ $predicate_j,$ $object_j \rangle$ or  $\langle resource_j,$ $propertie_j,$ $statement_j \rangle$. 
For instance, we have information about a rescue point like ``\textit{a rescue point located at 17 Winston Churchill Street, Compiègne. This rescue point has 100 people who need a transit to a rescue place}'', the set of triplets for this information is expresses as 
$P \: = \: \{\langle RescuePoint\_01,\:rdf:type,\:cmo:RescuePoint \rangle, \: \langle RescuePoint\_01, \: has\_Total\_People,\\ \: 100 \rangle,\: \langle RescuePoint\_01, \: has\_Address, \: 17\_Winston\_\\Churchill\_Street\_Compi\grave{e}gne \rangle \}$.
While, shelters represent a public safe place that is arranged and managed by a Municipal Council. Similar to the rescue points, a shelter $S$ can be defined as a set of triplets as follows:\vspace{-0,2cm}
\begin{equation}
	S = \{a^S_1, a^S_2, ..., a^S_m\}
\end{equation}
where $a^S_t$ is the triplet $a^S_t$ $=\langle subject_t,$ $predicate_t,$ $object_t \rangle$ or  $\langle resource_t,$ $propertie_t,$ $statement_t \rangle$. For example, the Municipal Council of Compiègne city organizes a shelter at Rose Gymnasium with capacity of 200 people, this information can be represented as
$S \: = \: \{ \langle Shelter\_01, \: rdf:type, \: cmo:Shelter \rangle, \: \langle Shelter\_\\01, \: has\_capacity, \: 200 \rangle, \:\langle Shelter\_01, \: has\_Address,\\ \:Rose\_Gymnasium\_Compi\grave{e}gne \rangle \:\}$ . 


Finally, the populations can be distinguished into affected populations and human resources. The affected populations are the vulnerable populations in the crisis, and they need to move to a shelter. While the human resources can be drivers who use their vehicle to participate in evacuation activities. In general,  the representation of a person by ontology is useful in the gathering of the human resources, and information about affected people in the pre-stage and post-stage of the crisis.

\subsection{Constraint-based recommender system} \label{sec:constraint_based}

\vspace{-0.2cm}
In this section, we present in detail the constraint-based recommender system for crisis management simulations. As we introduce in Section~\ref{sec:recommender_system}, constraint-based recommender systems generate recommendations by identifying items that satisfy a set of predefined explicit constraints. The goal of our system is to recommend pairs of civil/volunteer driver/vehicle for each rescue point such that the vehicles distributed to each rescue point have enough places to evacuate the population at each rescue point while minimizing the time needed for arriving at the rescue points. 

Suppose that there are $x$ rescue points:  $\mathcal{P} = \{P_1, P_2, P_3, \dots, P_x\}$, the number of people who can move by themselves in rescue point $P$ is $NB_{P}$ and the number of people who cannot move by themselves (injured or disabled) is $NB_{P}^{DIS}$. For people who cannot move, we assume that they need more space and we empirically set the number of seats as 2. Suppose that there are $y$ usable civil/volunteer vehicles: $\mathcal{R} = \{R_1,R_2,R_3,\dots,R_y\}$, the seats available on each civil/volunteer vehicle is $PA_{R}$. To represent the distribution of each civil/volunteer vehicle, we construct a distribution matrix: $M \in \mathbb{R} ^ {x \times y}$ where $M_{u,v} \in \{0,1\}$, $M_{u,v} = 1$ means that civil/volunteer vehicle $R_v$ is distributed to rescue point $P_u$,  $M_{u,v} = 0$ means that civil/volunteer vehicle $R_u$ is not distributed to rescue point $P_v$. Basically, one civil/volunteer vehicle should be distributed to at most one rescue point. 

Formally, the constraints applied in the recommender system are the following:

 For any $v \in \{1,2,3,\dots, y\}$\vspace{-0,3cm}
 \begin{equation} \label{equa:constraint1}
    \sum\limits_{u=1}^{u=x}M_{u,v} \leq 1
\vspace{-0.4cm}
\end{equation}

For any $u \in \{1,2,3,\dots, x\}$\vspace{-0,3cm}
 \begin{equation} \label{equa:constraint2}
    \sum\limits_{v=1}^{v=y}M_{u,v} * PA_{R_v} \geq NB_{P_u} + 2 * NB_{P_u}^{DIS}
    \vspace{-0.4cm}
\end{equation}

Equation~\ref{equa:constraint1} ensures that each civil/volunteer vehicle is distributed to at most one rescue point and Equation~\ref{equa:constraint2} makes sure that the sum of seats available  on the civil/volunteer vehicles distributed to each rescue point is larger than the number of people to be evacuated at that point. 

We denote the time needed for civil/volunteer vehicle $R$ to arrive at rescue point $P$ as $T_{R-P}$, and the set of civil/volunteer vehicles distributed to rescue point $P$ is $R_{P}$. Therefore, the goal is to minimize the following:\vspace{-0,5cm}
 \begin{equation} \label{equa:goal}
    min \sum\limits_{v=1}^{v=y} \sum\limits_{R \in R_{P_v}} T_{R-P_v}
    \vspace{-0.2cm}
\end{equation}

That being said, when several solutions are available, our algorithm returns the one that uses fewer vehicles to reduce the total time needed and the risk of traffic jams. We compute $T_{CV-RP}$ by \emph{OSMNX} \cite{boeing2017osmnx}, a python package that allows downloading geospatial data from the \emph{OpenStreetMap}. We have tried \emph{Google Maps API}, and it turned out that using \emph{Google Maps API} took a longer time to calculate the time between two points than using \emph{OpenStreetMap}, therefore, we adopt \emph{OpenStreetMap}.


\section{Prototype} \label{sec:prototype}
\vspace{-0.2cm}
Following the design science search guidelines proposed by  \cite{hevner2004design}: ``Design-science research must produce a viable artifact in the form of a construct, a model, a method, or an instantiation", in this section, we present a prototype of a system that helps to recommend civil/volunteer resources (pairs of civil/volunteer driver/vehicle) during a crisis. We will first present the architecture of the system that implements the constraint-based recommender system in Section~\ref{sec:constraint_based}, how the system works and then we describe a scenario to illustrate the utility of our system.

\subsection{System architecture}
\vspace{-0.2cm}
\begin{figure}
\vspace{-0.2cm}
	\centering
	\includegraphics[width=0.4\textwidth]{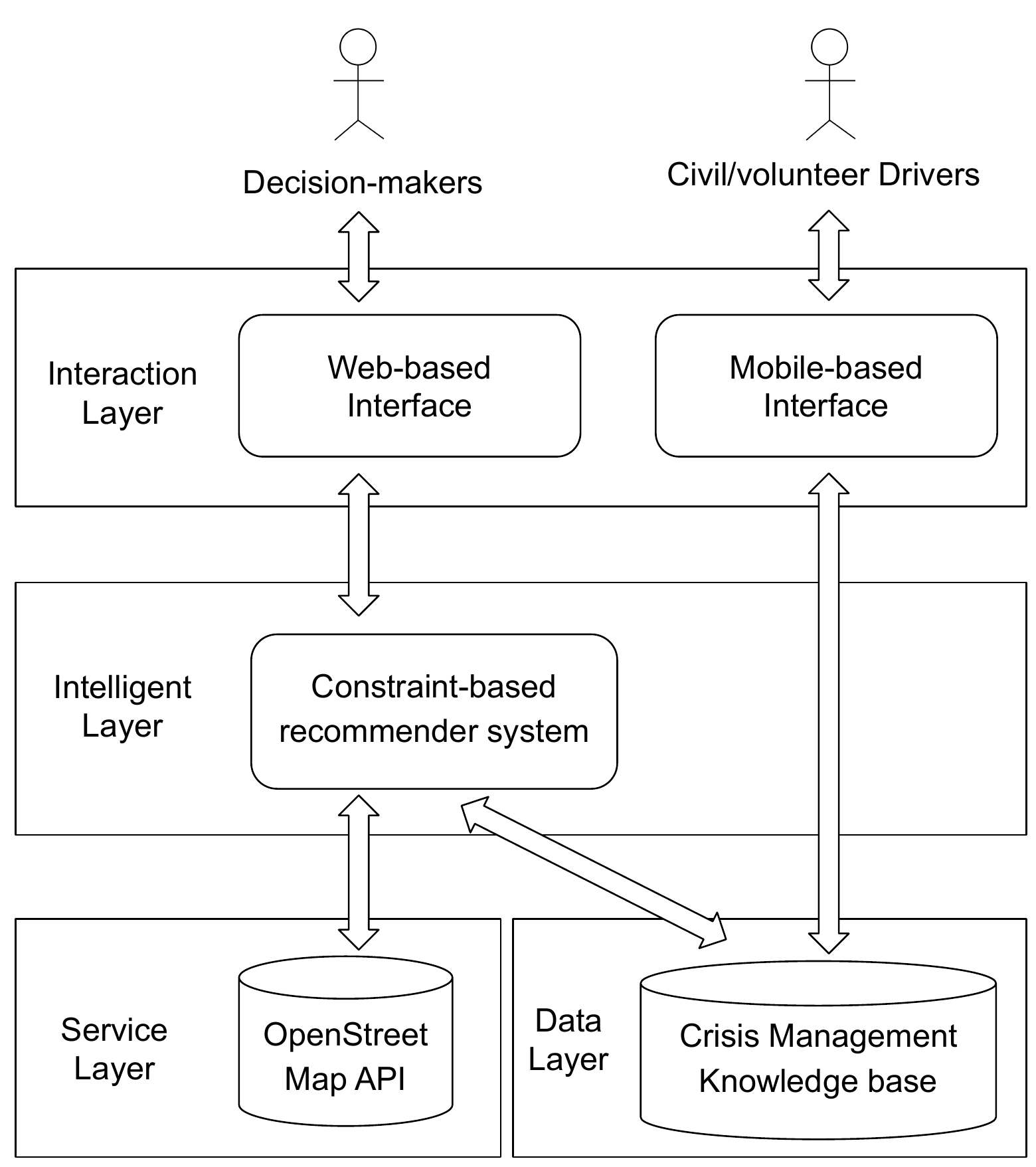}
	\caption{The architecture of our system}
	\label{fig:schema}
    \vspace{-0.4cm}
\end{figure}

Figure~\ref{fig:schema} shows the architecture of our system, which is composed of four layers: \emph{Interaction Layer}, \emph{Intelligent Layer}, \emph{Service Layer} and \emph{Data Layer}. 

\noindent\textbf{\emph{Interaction Layer:}} The interactions in the \emph{interaction layer} are two-fold, on the one hand, the mobile-based interface (e.g. a mobile app) collects civil/volunteer drivers' disponibility, which helps the system identify the available civil/volunteer drivers (with their available vehicles) during a crisis. Ideally, the app informs the volunteers once the decision-makers have made a request for help from the system; on the other hand, the web-based interface is designed for decision-makers to interact with the system. Decision-makers specify the information of each rescue point: the address of the rescue point, the number of people (people who can move by themselves and people who cannot move by themselves) to be evacuated at each point, and the priority level of each point. The system returns a list of recommendations (pairs of civil/volunteer driver/vehicle) for each rescue point. Note that the priority level defined by decision-makers decides the presentation order for each rescue point. The higher the priority level, the topper the recommendations for the rescue point will be presented, we will illustrate this with a scenario study in Section~\ref{sec:scenario}.

\noindent\textbf{\emph{Intelligent Layer:}} This layer is the core of the system that computes the list of recommendations satisfying the constraints described by Equations~\ref{equa:constraint1}~and~\ref{equa:constraint2}. The recommendations are displayed to decision-makers through the web-based interface.

\noindent\textbf{\emph{Service Layer:}} The objective of this layer is to compute the coordinates (longitude and latitude) of each rescue point given the address of each point. With the help of \emph{OpenStreetMAP} and the python package \emph{OSMNX}, the time $T_{R - P}$ for each distance between the pair of civil/volunteer driver/vehicle $R$  and the rescue point $P$ is computed. The time computed is used to filter pairs of civil/volunteer driver/vehicle, see Equation~\ref{equa:goal}.

\noindent\textbf{\emph{Data Layer:}} This layer contains an ontology-supported knowledge base for the crisis management domain. It models and stores all the necessary information and data related to the rescue points, affected people, and pairs of driver/vehicle, for more detail, see Section~\ref{sec:construction_of_ontology}.

It can be seen that our system follows a decoupling architecture, each layer is tightly connected. The advantage of such architecture is the reusability of each component. For example, when a new algorithm is developed in the \emph{Intelligent Layer}, the components in other layers can be reused, which saves engineering costs.

\subsection{Scenario} \label{sec:scenario}
    \vspace{-0.3cm}
\begin{table}[]
    \centering
    \footnotesize{
    \begin{tabular}{|M{0.14\textwidth}|M{0.07\textwidth}|M{0.07\textwidth}|M{0.08\textwidth}|}
        \hline
         \centering Civil/volunteer vehicle type & \centering Quantity &  \centering Total seats & Total lying places  \\\hline
         Minibus & 6 & 95 & 0\\\hline
         Minivan & 5 & 18  & 18 \\\hline
         Van & 5 & 45 & 0\\\hline
         Campervan & 1 & 4 & 6\\\hline
         SUV, crossover & 20 & 80 & 0\\\hline
         Berline & 15 & 60 & 0\\\hline
    \end{tabular}
    }
    \caption{Civil/volunteer vehicle resources can be mobilized}
    \label{table_01}
        \vspace{-0.6cm}
\end{table}

\begin{table}[]
    \centering
    \footnotesize{
    \begin{tabular}{|c|c|c|}
        \hline
        Shelter Name  & Capacity\\\hline
         Rose Gymnasium & 320 \\\hline
         Gaëtan Denain Gymnasium & 120 \\\hline
         Tainturier Gymnasium & 240 \\\hline
    \end{tabular}
    }
    \caption{The list of shelters}
    \label{table_02}
    \vspace{-0.4cm}
\end{table}

\begin{figure}[h]
		\centering
		\frame{\includegraphics[width=0.47\textwidth]{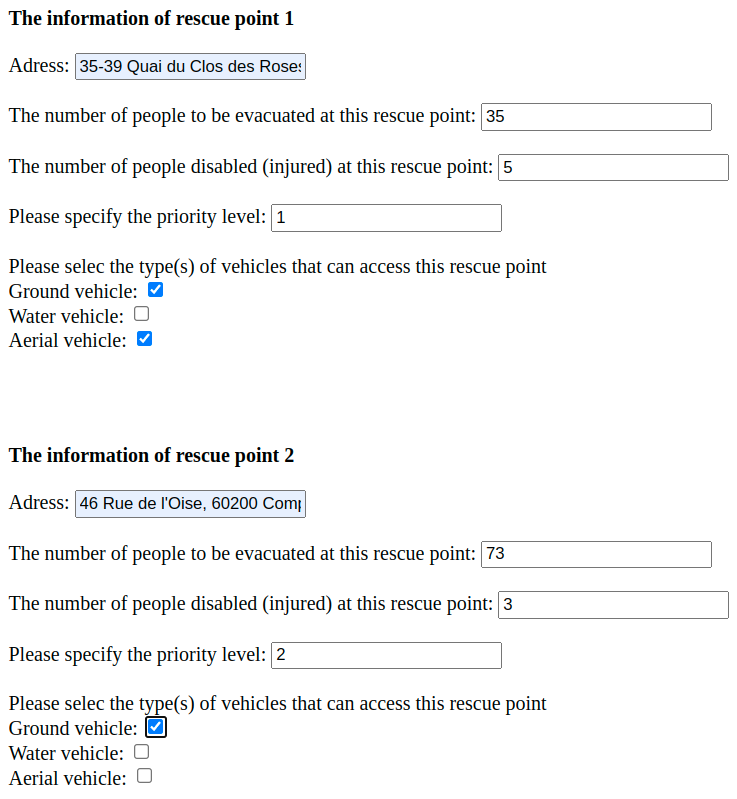}}
		\caption{The interface used to enter necessary information for each rescue point}
		\label{fig_03}
    \vspace{-0.3cm}
	\end{figure}

In this section, we present a scenario where our proposed system can be applied.
At the \textit{Compiègne} city,  a flood crisis is identified, and it is notified to all the population involved. The person in charge at the Municipal Council initiates necessary evacuation activities in order to move vulnerable people by the flood to a secure place. According to the current situation, they have received information about two areas that are being gradually submerged by this flood, including the number of people and disabled people, and priority level in each area (see Figure \ref{fig_03}). 
	
	The Municipal Council previously has collected and listed pairs of civil/volunteer driver/vehicle that can be used and mobilized in cases of emergencies like a current situation. Thus, they have about 50 civil/volunteer vehicles with different types and their drivers that can be used. The list of civil/volunteer vehicle resources is categorized by the type of the vehicle that is shown in Table \ref{table_01}. Furthermore, they also expropriate and prepare three shelters, which are  the gymnasiums for people staying during the crisis (see Table \ref{table_02}). Based on our knowledge base model, information of these pairs of civil/volunteer driver/vehicle resources and these shelters have been organized and entered into the system by a mobile-based application. 
	
	Back to the flood situation,  people in the two previously identified areas need to be moved to a secure place as soon as possible. Therefore, two rescue points corresponding to two dangerous areas are identified along with information about the address and the number of people, the number of disabled people and priority level. The question raised in this situation is how precisely and optimally allocate resources in terms of civil/volunteer driver/vehicle for each rescue point in order to move vulnerable people in the shortest time to a shelter.
	
\begin{figure*}[h]
		\centering
		\frame{\includegraphics[width=0.86\textwidth]{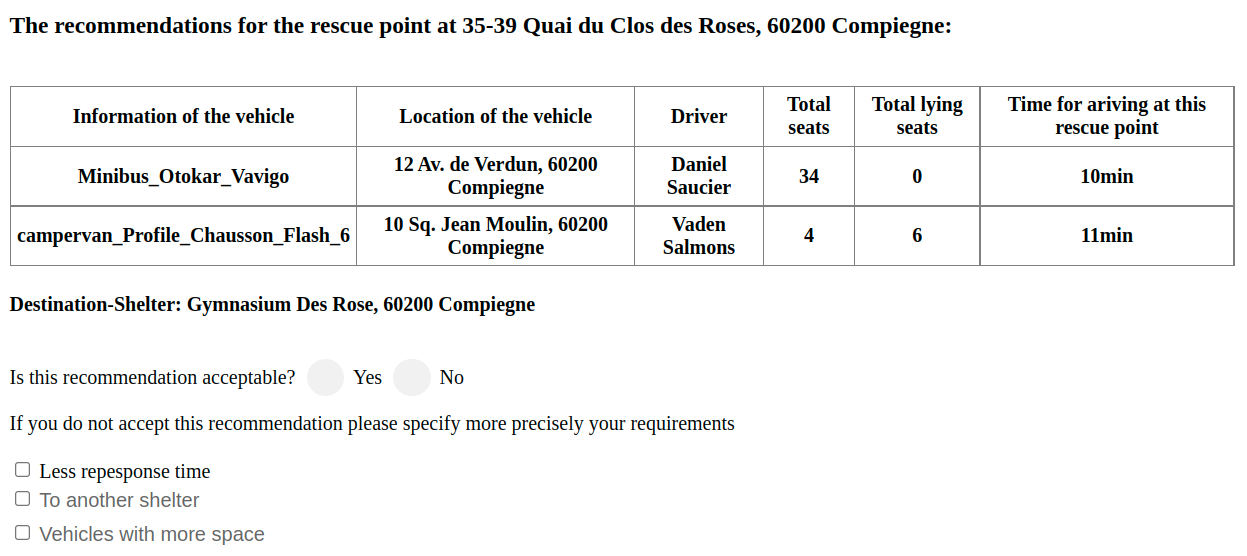}}
		\caption{The result of the driver/vehicle allocation for the secure point}
		\label{fig_04}
		    \vspace{-0.3cm}
	\end{figure*}

The answer to this question can be implemented by our system. Firstly, the system receives information about the requirements of each rescue point with a web-based interface (from decision-makers). More specifically, the interface represents the required information of two rescue points as shown in Figure~\ref{fig_03}.   Secondly, the system collects the information about rescue points and begins gathering the information about available driver/vehicle resources in the knowledge base data model, as well as the information about estimated time and distance to moving between vehicle and rescue point by using OpenStreetMap data, which has to be minimized, see Equation~\ref{equa:goal}. Finally, the system calculates and recommends an optimal solution with the suggestion of a list of pairs of civil/volunteer driver/vehicle for each rescue point as in Figure \ref{fig_04}. The decision-makers will decide whether the returned civil/volunteer drivers/vehicles are appropriate or not. If the decision-maker thinks that it is appropriate, then the civil/volunteer drivers/vehicles are sent out; if not, the decision-maker can revise the requirements, for example, less time for arriving or more places in the vehicles.

\subsection{Discussion}
\vspace{-0.2cm}

This work aims at constructing a knowledge base in the crisis management domain and implementing a constraint-based recommender system by recommending relevant solutions for allocating pairs of civil/volunteer driver/vehicle resources with the help of an ontology-supported knowledge base. 


We point out some limits of this work: (1) During a crisis, for example, a flood, the situation may change from time to time (in evolution), and the depth of water can influence the choice of vehicle. When vehicles on the water have to be used, the route plan is one challenge to be resolved since the \emph{OpenStreetMap} cannot make a route plan when the roads are submerged by water. A foreseeable solution would be to use a boat to transfer people to a place where a ground vehicle is accessible and then use a ground vehicle to transfer people to the shelter. (2) When the number of civil/volunteer drivers/vehicles is limited, they have to be reused, which means that after dropping off the people evacuated at the shelter, the drivers/vehicles can be distributed to other rescue points. In this case, our system becomes dynamic. Clearly, the \emph{Data Layer} (see Figure~\ref{fig:schema}) has to be adapted. Due to the modular structure of our system, other components stay unchanged.

We now point out the traits of this system. The knowledge base supported by an ontology allows for efficient and standardized organization and collection of information. The approach to the development of a recommender system using a knowledge base and constraints helps to create flexibility in defining rules and constraints in different situations and contexts. The scenario proves the utility of our proposed system in crisis management that can be deployed in a computer-simulated environment and maybe the real world. The modular structure of our system is another important advantage, each component can be reused when other components are modified, which helps to save engineering efforts. Therefore, we believe that although the functions of our system are limited for the moment, the decoupling architecture of our system facilitates the updates and development of more functions in the future.

\section{Conclusion and perspectives}
\vspace{-0.4cm}
We now conclude and propose some avenues for future work. In this paper, we propose a recommender system that can help decision-makers distribute pairs of civil/volunteer driver/vehicle when such public resources could not be enough. Our system follows a modular structure that is composed of four layers. The ontology-supported \emph{Data Layer} structures and stores the necessary data; the \emph{Service Layer} allows the calculation of time and distance between two geographic points by applying \emph{OpenStreetMap}; the \emph{Intelligent Layer} computes a list of recommendations for each rescue point; the \emph{Interaction Layer} facilitates the interactions between decision-makers and the recommender system, and allows to identify the available pairs of driver/vehicle during a crisis. Each layer is tightly connected and such a structure promotes the reusability of our system. 

Here are some foreseeable avenues for future work: (1) We plan to enrich our ontology to deal with the complicated situation during a crisis that is in evolution. (2) We plan to add more constraints to better model the real situation of crisis. For example, during a flood, we can add the depth of water to determine the types of boats usable. The experiences of drivers are another important factor to be considered: the number of years drivers have obtained driver's license, the rescue training conducted before. 
(3) The former two plans lead to our third plan: building a dynamic system that enables the reuse of resources (drivers and vehicles) in a real-time way.
(4) We intend to test and integrate our proposed system into a complete agent-based simulation system such as the work in \cite{laatabi2022coupling} in order to obtain more experiments and evaluations in different scenarios of crisis management, and to explore the  socio-technical aspects of our system \cite{land2000evaluation}.

\section*{Acknowledgement}
   \vspace{-0.3cm}
This work was funded by the French Research Agency (ANR) and by the company Vivocaz under the project France Relance - preservation of R\&D employment (ANR-21-PRRD-0072-01).
\bibliographystyle{ieeetr}
\bibliography{biblio}

\end{document}